\documentclass[showpacs, showkeys,twocolumn,amsmath,amssymb,prb]{revtex4}

\usepackage{graphicx}
\usepackage{dcolumn}
\usepackage{bm, url}
\begin{document}
\newcommand{\ud}{d}
\newcommand{\hs}{\hspace*{0.5cm}}

\newtheorem{1}{Proposition}
\newtheorem{2}{Theorem}
\newtheorem{A2}[2]{Theorem}

\title{The incomplete beta function law for parallel tempering sampling of classical canonical systems}

\author{Cristian Predescu}
\affiliation{Department of Chemistry,  Brown University, Providence, Rhode Island 02912}
\author{Mihaela Predescu}
\affiliation{Department of Mathematics, Bentley College, Waltham, Massachusetts 02452}
\author{Cristian V. Ciobanu }
\affiliation{Division of Engineering, Brown University, Providence, Rhode Island 02912}
\date{\today}

\begin{abstract} 
We show that the acceptance probability for  swaps in the parallel tempering Monte Carlo method for classical canonical systems is given by a universal function  that depends on the average statistical fluctuations of the potential and on the ratio of the temperatures. The law, called the incomplete beta function law, is valid in the limit that the two temperatures involved in swaps are close to one another. An empirical version of the law, which involves the heat capacity of the system, is developed and tested on a Lennard-Jones cluster. We argue that the best initial guess for the distribution of intermediate temperatures for parallel tempering is a geometric progression and we also propose a technique for the computation of optimal temperature schedules. Finally, we demonstrate that the swap efficiency of the parallel tempering method for condensed-phase systems decreases naturally to zero at least as fast as the inverse square root of the dimensionality of the physical system. 
\end{abstract}

\pacs{05.10.Ln, 02.70.Tt}
\keywords{Monte Carlo, parallel tempering, replica exchange, acceptance probability}

\maketitle

\section{Introduction}

Ergodic Monte Carlo simulations of large dimensional systems having complicated topologies with many disconnected local minima are difficult computational tasks, though indispensable for many physical applications.\cite{Swe86, Tes96, Han97, Wo99, Fal99, Yan99, Nei00, Cal00, Yan00, Sug00, Yan01, Cal01, Bed01, Gro01, Ish01, Bun01, Fal02, Fuk02} Among the various methods dealing with such problems, the parallel tempering method\cite{Gey95, Huk96} is one of the most successful, especially given the simplicity of the idea and the ease of implementation. For sure, the idea of coupling two independent Markov chains characterized by different parameters in order to ensure transfer of information from one to the other has a long history. In physical sciences, coupling strategies have been employed for the development of specialized sampling techniques such as replica Monte Carlo sampling of spin glasses,\cite{Swe86} jump-walking,\cite{Fra90, Whi01} and simulated tempering,\cite{Lyu92, Mar92} to give a few examples. How this coupling must be performed in the simplest, most general, and most efficient way is, however, a quite difficult problem. 

The parallel tempering method, as we utilize it in this article, addresses the question of coupling independent Monte Carlo chains that sample classical Boltzmann distributions for different temperatures and which are usually generated by the Metropolis \emph{et al} algorithm.\cite{Met53, Kal86} The method has been formalized seemingly independently by Geyer and Thompson\cite{Gey95} as well as by Hukushima and Nemoto.\cite{Huk96} Of course, it is not necessary that the distributions involved in swaps differ through their temperature. For instance, the controlling parameter may be the chemical potential, as in the hyperparallel tempering method,\cite{Yan99} a delocalization parameter, as in the q-jumping Monte Carlo method,\cite{And97}  or suitable modifications of the potential, as in the Hamiltonian replica exchange method.\cite{Sug00, Fuk02} 

 In parallel tempering, swaps involving two temperatures $\beta$ and $\beta'$ are attempted from time to time in a cyclic or random fashion and accepted with the conditional probability
\begin{equation}
\label{01}
\min\left\{1, e^{(\beta' - \beta)[V(\textbf{x}')-V(\textbf{x})]}\right\},
\end{equation}
where $V(\textbf{x})$ is the potential of the physical system under consideration. This acceptance rule ensures that the detailed balance condition\cite{Kal86} is satisfied and that the equilibrium distributions are the Boltzmann distributions. As Eq.~(\ref{01}) suggests, the efficiency of the temperature swaps depends on the difference between the inverse temperatures $\beta$ and $\beta'$. In order to maintain high acceptance ratios, only swaps between neighboring temperatures in a given schedule  $\beta_{min}=\beta_1 < \beta_2 < \cdots < \beta_{N} = \beta_{max}$ are attempted. An optimal schedule of temperatures has the property that the acceptance ratios between neighboring temperatures are equal to some predetermined value $p$, value that is usually greater or equal to $0.5$. The determination of the optimal schedule is complicated by the fact that the distributions of the coordinates $\textbf{x}$ and $\textbf{x}'$ are also temperature dependent (they are, of course, the Boltzmann distributions at the temperatures $\beta$ and $\beta'$, respectively). 

In this article, we attempt to answer the following important question: What are the main properties of the physical system that control the acceptance ratio in the limit that the difference $\beta' - \beta$ is small? The answer to this question allows for a better understanding of the applicability as well as the limitations of the parallel tempering method. In addition, it allows for the development of optimal temperature schedules in a way that seems  more direct and easier to implement than other adaptive strategies.\cite{Huk96}

 In Section~II, we demonstrate in a rigorous mathematical fashion that the acceptance probabilities for parallel tempering swaps are controlled, within an $O(|\beta'-\beta|^3)$ error, by the \emph{ratio} of the two temperatures involved in swaps and by the \emph{average potential fluctuations} of the system, at the inverse temperature $\bar{\beta}= (\beta + \beta')/2$. The acceptance probabilities are well approximated by the so-called incomplete beta function law, which has the additional property that it is exact for harmonic oscillators.  Under the assumption that the relation between the average fluctuations and the average square fluctuations of the potential is roughly the one for harmonic oscillators, we develop an empirical version of the incomplete beta function law, version that connects the acceptance probabilities for parallel tempering swaps with the temperature ratios and the heat capacity of the system. 
 
In Section~III.B, we show how the incomplete beta function laws can be employed for the determination of optimal temperature schedules. We also explain the empirical observation that a geometric progression is the best schedule for systems and ranges of temperatures for which the heat capacity is almost constant.\cite{Sug00} In Section~III.C, we demonstrate rigorously that the efficiency of the parallel tempering method for harmonic oscillators decreases naturally to zero at least as fast as the inverse square root of the dimensionality of the physical system. We then argue that  the loss in efficiency is even greater for condensed-phase systems (both solids and liquids). This result seems to be in contradiction with the findings of Kofke,\cite{Kof02} who suggested that such a curse of dimension does not appear for parallel tempering. However, the result is  in agreement with the explanation of Fukunishi, Watanabe, and Takada.\cite{Fuk02} It is for this reason that we insist that our findings be proven in a mathematically rigorous way. 

The rigorous form of the incomplete beta function law involves the average potential fluctuations at certain temperatures. An evaluation of this property by Monte Carlo simulations would require the computation of a double integral over the configuration space. For this reason, current Monte Carlo codes would have to be modified extensively in order to take advantage of the incomplete beta function law for the design of optimal temperature schedules. To circumvent this undesirable situation, we propose an empirical version of the incomplete beta function law, version that is derived under the assumption that the relation between the average fluctuations and the average square fluctuations of the potential is roughly the one for harmonic oscillators. In Section~IV, we illustrate the good applicability of the empirical law by performing a Monte Carlo simulation for a cluster made up of $13$ atoms of neon that interact through Lennard-Jones potentials.

\section{The incomplete beta function law}
Consider a $d$-dimensional physical system  described by the  potential $V(\textbf{x})$, which is assumed to be bounded from below. To simplify notation,  we may also assume that the global minimum of the potential is $0$, perhaps after addition  of a constant. Clearly, the addition of a constant does not change the acceptance probabilities for swaps. 

In the parallel tempering algorithm, swaps involving two temperatures $\beta$ and $\beta'$ occur with the conditional probability
\begin{eqnarray*}&&
\min\left\{1, e^{(\beta' - \beta)[V(\textbf{x}')-V(\textbf{x})]}\right\}.
\end{eqnarray*}
The joint probability distribution density of the points $\mathbf{x}$ and $\mathbf{x}'$ in an equilibrated system is given by the formula
\[
\frac{1}{Q(\beta)Q(\beta')}e^{-\beta V(\mathbf{x})}e^{-\beta' V(\mathbf{x}')}, 
\]
where
\[
Q(\beta) = \int_{\mathbb{R}^d} e^{-\beta V(\mathbf{x})} d \mathbf{x}
\]
is the configuration integral. 
It follows that the acceptance probability $Ac(\beta, \beta')$ for swaps between neighboring temperatures is given by the average
\begin{eqnarray}
\label{eq:1}
Ac(\beta,\beta')= \frac{1}{Q(\beta)Q(\beta')}\int_{\mathbb{R}^d}d \mathbf{x} \int_{\mathbb{R}^d}d \mathbf{x}'e^{-\beta V(\mathbf{x})}e^{-\beta' V(\mathbf{x}')}\nonumber \\ \times \min\left\{1, e^{(\beta' - \beta)[V(\mathbf{x}')-V(\mathbf{x})]}\right\}. \quad
\end{eqnarray}

By construction, $Ac(\beta, \beta')$ is symmetrical under exchange of variables. Without loss of generality, we may assume $\beta' \geq \beta$. Then,
\begin{eqnarray}
\label{eq:2}
\nonumber 
\min\left\{1, e^{(\beta' - \beta)[V(\mathbf{x}')-V(\mathbf{x})]}\right\} =
e^{(\beta' - \beta)\min\{0, V(\mathbf{x}')-V(\mathbf{x})\}} \\ =
e^{\frac{(\beta' - \beta)}{2} [V(\mathbf{x}')-V(\mathbf{x})]} e^{-\frac{(\beta' - \beta)}{2} |V(\mathbf{x}')-V(\mathbf{x})|}. \quad 
\end{eqnarray}
Replacing Eq.~(\ref{eq:2}) in Eq.~(\ref{eq:1}), we obtain
\begin{eqnarray}
\label{eq:3}
Ac(\beta,\beta')= \frac{1}{Q(\beta)Q(\beta')}\int_{\mathbb{R}^d}d \mathbf{x} \int_{\mathbb{R}^d}d \mathbf{x}'e^{-\bar{\beta} V(\mathbf{x})}e^{-\bar{\beta} V(\mathbf{x}')}\nonumber \\ \times \exp\left[-\frac{R-1}{R+1}\bar{\beta}\left|V(\mathbf{x}')-V(\mathbf{x})\right|\right]. \quad
\end{eqnarray}
In Eq.~(\ref{eq:3}), the variables $R$ and $\bar{\beta}$ are defined by the equations
\begin{equation}
\label{eq:4}
R = \frac{\beta'}{\beta} \quad \text{and} \quad \bar{\beta} = \frac{\beta + \beta'}{2},
\end{equation}
respectively. 
Due to the nature of the results we prove, it is more convenient to express the various formulas in terms of the new variables $R$ and $\bar{\beta}$. Because $\beta'\geq \beta$, we need only consider the case $R\geq 1$.

As announced in the introduction, we are interested in establishing asymptotic laws in the limit that $R \to 1$ for which the error is of order $O\left(|\beta'-\beta|^3\right)$ or, alternatively, $O\left(|R - 1|^3\right)$. At this point, let us see that the acceptance probability given by Eq.~(\ref{eq:3}) is alternatively given by the formula
\begin{equation}
\label{eq:5}
Ac(\beta,\beta') = \frac{Q(\bar{\beta})^2}{Q(\beta)Q(\beta')} \left\langle \exp\left[-\frac{R-1}{R+1}\bar{\beta}\left|U'-U\right|\right] \right\rangle_{\bar{\beta}},
\end{equation}
where, in general, $\left\langle f(U,U') \right\rangle_{\beta}$ denotes the statistical average
\begin{equation}
\label{eq:6}
 \frac{1}{Q(\beta)^2}\int_0^\infty\int_0^\infty  e^{-\beta U}e^{-\beta U'}  \Omega(U)\Omega(U') f(U,U')dU dU'.
\end{equation}
In Eq.~(\ref{eq:6}), $\Omega(U)$ denotes the density of states.

In these conditions, the following proposition holds true.
\begin{1}
We have 
\begin{equation}
\label{eq:7}
Ac(\beta, \beta') = 1 - \frac{R-1}{R+1}M\left(\bar{\beta}\right) + O\left(|R - 1|^3\right),
\end{equation}
where 
\begin{eqnarray}
\label{eq:8}
M(\bar{\beta})= \bar{\beta} \left\langle \left|U' - U\right| \right\rangle_{\bar{\beta}}.
\end{eqnarray}
\end{1}

\emph{Proof.}
A Taylor expansion of the exponential function to the third order and the identity $\left\langle U' - U\right\rangle_{\bar{\beta}} = 0$ imply
\begin{eqnarray*}&&
\frac{Q(\beta)Q(\beta')}{Q(\bar{\beta})^2} = \left\langle \exp\left[- \frac{R-1}{R+1}\bar{\beta}(U'-U)\right]  \right\rangle_{\bar{\beta}}   
 \\ && = 1 + \frac{1}{2}\left(\frac{R-1}{R+1}\right)^2 \bar{\beta}^2\left\langle |U' - U|^2\right\rangle_{\bar{\beta}} + O(|R-1|^3). 
\end{eqnarray*}
Therefore,
\begin{eqnarray}
\label{eq:9}\nonumber
\frac{Q(\bar{\beta})^2}{Q(\beta)Q(\beta')}  = 1 - \frac{1}{2}\left(\frac{R-1}{R+1}\right)^2 \bar{\beta}^2\left\langle |U' - U|^2\right\rangle_{\bar{\beta}}\\ + O(|R-1|^3). 
\end{eqnarray}
On the other hand,
\begin{eqnarray} 
\label{eq:10}
\nonumber
 \left\langle \exp\left[-\frac{R-1}{R+1}\bar{\beta}\left|U'-U\right|\right] \right\rangle_{\bar{\beta}} = 1 - \frac{R-1}{R+1}M(\bar{\beta}) \\ +
 \frac{1}{2}\left(\frac{R-1}{R+1}\right)^2 \bar{\beta}^2\left\langle |U' - U|^2\right\rangle_{\bar{\beta}} + O(|R-1|^3).
\end{eqnarray}
Eq.~(\ref{eq:9}) and (\ref{eq:10}) imply
\begin{eqnarray*}
\frac{Q(\bar{\beta})^2}{Q(\beta)Q(\beta')} \left\langle \exp\left[-\frac{R-1}{R+1}\bar{\beta}\left|U'-U\right|\right] \right\rangle_{\bar{\beta}}\\ = 1 - \frac{R-1}{R+1}M(\bar{\beta}) + O(|R-1|^3)
\end{eqnarray*}
and the proof is concluded. \hspace{\stretch{1}} $\Box$

Proposition~1, while a powerful asymptotic result, has the disadvantage that it may produce negative numbers for the acceptance probability in actual simulations. However, we can repair this in very straightforward fashion. Notice that the fact that $\Omega(U)$ does not depend upon the temperature is \emph{not crucial} for the proof of Proposition~1. Thus, given any other well-behaved density of states $\Omega'[U, d(\beta)]$ depending perhaps on the inverse temperature through an adjustable parameter $d(\beta)$ and such that
\begin{equation}
\label{eq:11}
\bar{\beta}\left\langle |U-U'|\right\rangle'_{\bar{\beta}} = M\left(\bar{\beta}\right),
\end{equation}
we still have
\begin{equation}
\label{eq:12}
Ac'(\beta, \beta') =  1 - \frac{R-1}{R+1}M\left(\bar{\beta}\right) + O\left(|R - 1|^3\right).
\end{equation}
In Eqs.~(\ref{eq:11}) and (\ref{eq:12}), the prime sign denotes the respective averages or quantities obtained from their definition if $\Omega(U)$ is replaced by $\Omega'[U,d(\beta)]$.
From Eqs.~(\ref{eq:7}) and (\ref{eq:12}), we deduce 
\begin{equation}
\label{eq:13}
Ac'(\beta, \beta') =  Ac(\beta, \beta') + O\left(|R - 1|^3\right)
\end{equation}
for all $\Omega'[U,d(\beta)]$ that satisfy Eq.~(\ref{eq:11}).  This simple observation allows us to construct approximations $Ac'(\beta, \beta')$ of order $O\left(|R - 1|^3\right)$ for the acceptance probability $Ac(\beta, \beta')$ by considering a special functional dependence for the density of states $\Omega'[U,d(\beta)]$, such that the resulting approximation is exact for a certain class of physical systems. 

We take this class to be the harmonic oscillators, for which $\Omega'(U, d) = (2\pi)^{d/2}\Gamma(d/2)^{-1} U^{d/2 - 1}$. We prove in Appendix~I that for any  $d$-dimensional harmonic oscillator,
\begin{equation}
\label{eq:14}
Ac_H(\beta,\beta') = \frac{2}{B(d/2,d/2)} \int_0^{\frac{1}{1 + R}} \theta^{d/2 - 1}(1 - \theta)^{d/2 -1}d \theta
\end{equation}
and
\begin{equation}
\label{eq:15}
M_H(\bar{\beta}) = 2^{2-d}B(d/2,d/2)^{-1}.
\end{equation}
Here, $B(d/2,d/2)$ denotes the respective Euler's beta function. 

For an harmonic oscillator, $d$ is the dimension. For general systems, $d = d(\beta)$ is just a fitting parameter chosen such that Eq.~(\ref{eq:11}) is satisfied. In fact, Eq.~(\ref{eq:11}), which in our case reads
\[
2^{2-d}B(d/2,d/2)^{-1} = M(\bar{\beta}),
\]
has always a unique solution because $2^{2-d}B(d/2,d/2)^{-1}$ increases strictly  from $0$ to $+\infty$, as $d$ also increases from $0$ to $+\infty$. The following theorem is an immediate consequence of Eq.~(\ref{eq:13}) and of the discussion above.
\begin{2}[Incomplete beta function law]
Consider an arbitrary thermodynamic system  for which $M(\beta)$ is given as a function of temperature. Let $d(\bar{\beta})$ be the unique solution of the equation
\begin{equation}
\label{eq:16}
2^{2-d}B(d/2,d/2)^{-1} = M(\bar{\beta}).
\end{equation}
Then,
\begin{eqnarray}
\label{eq:17}
\nonumber
Ac(\beta,\beta') = \frac{2}{B\left[d(\bar{\beta})/2,d(\bar{\beta})/2\right]} \int_0^{1/(1 + R)} \theta^{d(\bar{\beta})/2 - 1}\\ \times (1 - \theta)^{d(\bar{\beta})/2 -1}d \theta + O\left(|R-1|^3\right), \quad
\end{eqnarray}
with the error term canceling for harmonic oscillators. 
\end{2}

\section{Consequences of the incomplete beta function law}
There are several important  consequences of Theorem~1. The first one concerns the development of an empirical law connecting the acceptance probabilities  with the heat capacity and the ratio of the temperatures involved in the parallel tempering swap. The second one regards  the optimal distribution of temperatures in parallel tempering Monte Carlo simulations. Yet a third one is the statement that the efficiency of the swaps between neighboring temperatures decreases naturally  as the inverse square root of the dimensionality of the system. We analyze these consequences in some detail in the remainder of the section. 

\subsection{The empirical incomplete beta function law}
The property $M(\bar{\beta})$ is not usually determined in simulations, nor is it measured in experiments. It is therefore necessary to relate it to other thermodynamic properties, more precisely to the heat capacity. In addition, it is desirable to develop a version of the incomplete beta function law involving the heat capacity rather then $M(\bar{\beta})$, even if the law has an empirical validity only.

 The quantity 
\[M(\bar{\beta})= \bar{\beta} \left\langle \left|U' - U\right| \right\rangle_{\bar{\beta}}\]
measures the statistical fluctuations of the potential $V(\mathbf{x})$ and is connected to the heat capacity of the system. More precisely, the Cauchy-Schwartz inequality says that
\[
\bar{\beta}^2 \left\langle \left|U' - U\right| \right\rangle_{\bar{\beta}}^2 \leq \bar{\beta}^2 \left\langle \left|U' - U\right|^2 \right\rangle_{\bar{\beta}}.
\]
However, 
\begin{eqnarray*}&&
\bar{\beta}^2\left\langle \left|U' - U\right|^2 \right\rangle_{\bar{\beta}} = \bar{\beta}^2 \frac{1}{Q(\beta)^2}\int_0^\infty\int_0^\infty  e^{-\beta U}e^{-\beta U'} \\ && \Omega(U)\Omega(U') (U' - U)^2dU dU'= 2\bar{\beta}^2 \Bigg\{ \frac{1}{Q(\beta)}\int_0^\infty   e^{-\beta U} \\ && \times \Omega(U)U^2 dU  - \frac{1}{Q(\beta)^2}\left[\int_0^\infty e^{-\beta U} \Omega(U)U dU\right]^2\Bigg\}.
\end{eqnarray*}
The last term in the equation above is twice the potential contribution $C_V(\bar{\beta})$ to the total heat capacity of the system. (In this article, the heat capacity is always expressed in units of the Boltzmann constant $k_B$.)  The total heat capacity sums both the potential and the kinetic average square fluctuations and is given by the well-known formula
\begin{equation}
\label{eq:18}
C(\bar{\beta}) = C_V(\bar{\beta}) + d/2.
\end{equation}
Then, the identity
\begin{equation}
\label{eq:18a}
\bar{\beta}^2\left\langle \left|U' - U\right|^2 \right\rangle_{\bar{\beta}} = 2 C_V(\bar{\beta})
\end{equation}
implies the inequality
\begin{equation}
\label{eq:19}
M(\bar{\beta})^2 \leq 2C_V(\bar{\beta}).
\end{equation}

Eq.~(\ref{eq:19}) suggests that $M(\bar{\beta})^2$ is an extensive property of the physical system, property that  may be very large  for systems for which the heat capacity is also large. In fact, Sterling's formula shows that
\[M_H(\bar{\beta})^2 = \left[2^{2-d}B(d/2,d/2)^{-1}\right]^2 \approx \frac{2d}{\pi}
\]
for large dimensional harmonic oscillators. The relation has a linear scaling with the dimensionality of the system (that is, with the number of particles). This scaling appears also for the heat capacity of harmonic oscillators
\[
C_{H,V}(\bar{\beta})= d/2.
\]

For systems in condensed phase, for which an harmonic superposition is roughly a good approximation of the Boltzmann distribution, one may safely assume that the functional relationship between $M(\bar{\beta})^2$ and $C_V(\bar{\beta})$ is not very far from the one for the harmonic oscillator. Of course, this is always true in the low temperature limit. In this conditions, the solution $d(\bar{\beta})$ of the equation
\[
2^{2-d}B(d/2,d/2)^{-1} = M(\bar{\beta}).
\] 
is approximately $d(\bar{\beta}) = 2C_V(\bar{\beta})$. Replacing the last result in Eq.~(\ref{eq:17}), we obtain the following \emph{empirical incomplete beta function law}:
\begin{eqnarray}
\label{eq:20}
\nonumber
Ac(\beta,\beta') \approx \frac{2}{B\!\left(C_V(\bar{\beta}),C_V(\bar{\beta})\right)} \int_0^{1/(1+R)} \theta^{C_V(\bar{\beta}) - 1}\\ \times (1 - \theta)^{C_V(\bar{\beta}) -1}d \theta. \quad 
\end{eqnarray}
The good applicability of Eq.~(\ref{eq:20}) for realistic physical systems will be illustrated in Section~IV for the case of a  Lennard-Jones cluster. Eq.~(\ref{eq:20}) can be expressed in terms of the full heat capacity of the system with the help of Eq.~(\ref{eq:18}). The empirical incomplete beta function law is still exact for harmonic oscillators.

\subsection{On the optimal schedule of temperatures for the parallel tempering simulation}
It is an empirical observation\cite{Sug00} that the optimal schedule
(i.e., the schedule for which all the acceptance probabilities for swaps between
neighboring temperatures are equal) is given by a geometric progression
of temperatures,  if the heat capacity of the system is approximatively
constant for the range  $[\beta_{min},\beta_{max}]$. The incomplete beta function law gives a direct explanation of this phenomenon. As discussed in the previous section, the quantities $M(\bar{\beta})$ and $C_V(\bar{\beta})$ measure essentially the same information: average potential fluctuations and average potential square fluctuations, respectively. Thus, we expect that $M(\bar{\beta})$ is also constant in regions where $C_V(\bar{\beta})$ is. As shown by Eq.~(\ref{eq:17}), in these conditions,  the acceptance probabilities are a function of the ratio of the temperatures involved in swaps only.  Therefore, if the predetermined number of replicas is $N$ (with $N$ large enough so that Theorem~1 applies), then the optimal schedule is
\begin{equation}
\label{eq:21}
\beta_i = R^{i-1}\beta_{min},\quad 1 \leq i \leq N,
\end{equation}
where
\begin{equation}
\label{eq:22}
R = (\beta_{max}/\beta_{min})^{1/(N-1)}.
\end{equation}

If one trusts it, the empirical incomplete beta function law can be used to evaluate the optimal schedule for arbitrary systems in the following way. First, one computes a set of values for the potential part $C_V(\beta)$ of the heat capacity over the interval $[\beta_{min}, \beta_{\max}]$ using a geometric progression law. The results are then interpolated by cubic spline, for example. Only a rough estimate is necessary. Given a prescribed value $p$ for the acceptance probability  and given the inverse temperature $\beta_i$, one computes $\beta_{i+1}$ by solving the equation
\begin{eqnarray}
\label{eq:22a}\nonumber
\frac{2}{B\!\left(C_V(\bar{\beta}),C_V(\bar{\beta})\right)} \int_0^{(1 + \beta_{i+1}/\beta_i)} \theta^{C_V(\bar{\beta}) - 1}\\ \times (1 - \theta)^{C_V(\bar{\beta}) -1}d \theta = p,
\end{eqnarray}
where $\bar{\beta} = (\beta_i + \beta_{i+1})/2$. One starts with $\beta_1 = \beta_{min}$ and continues the procedure until the current inverse temperature becomes greater or equal to $\beta_{max}$. This way, one determines both the optimal distribution of temperatures and the minimal number of temperatures compatible with the prescribed acceptance probability $p$. To ensure the validity of the approximation furnished by Theorem~1, the value of $p$  should be large enough. In fact, values larger or equal to $0.5$ are necessary anyway in order to have good mixing between the Monte Carlo walkers running at neighboring temperatures. 

Sure enough, one may use the full incomplete beta function law to find the optimal schedule. However, the computation of $M(\bar{\beta})$ requires extensive changes to the existing codes. In addition, as illustrated by the example described in Section~IV, the empirical version of the incomplete beta function law may be sufficiently accurate for most systems of practical interest. The applicability of the law can also be tested during the computation of the heat capacity $C_V(\beta)$, by comparing the observed values for the acceptance probabilities with the ones predicted by the empirical incomplete beta function law. 

\subsection{Loss of efficiency with increasing dimension for the parallel tempering method} 

In this subsection, we show that  the minimum number of intermediate temperatures $N(d,p)$ that ensures an acceptance probability greater or equal to some preset value $p\in(0,1)$ for a $d$-dimensional system increases naturally at least as the square root of the dimensionality for condensed-phase systems (both solids and liquids). This observation was first made by Hukushima and Nemoto.\cite{Huk96} 

We begin with a rigorous mathematical analysis for harmonic oscillators.
For them, the optimal schedule is a geometric progression [because $M(\bar{\beta})$ is independent of temperature] and the minimum value of $N(d,p)$ is given by
\begin{equation}
\label{eq:23}
N_H(d,p) = \left[\frac{\ln(\beta_{max}/\beta_{min})}{\ln[R(d,p)]} \right] + 2,
\end{equation}
where $[x]$ is the integer part of $x$ and $R(d,p)$ is the solution of the equation
\[ \frac{2}{B(d/2,d/2)} \int_0^{\frac{1}{1 + R}} \theta^{d/2 - 1}(1 - \theta)^{d/2 -1}d \theta = p. \]
As shown by Theorem~2 of Appendix~2,
\[
\lim_{d \to \infty} \frac{2}{B(d/2,d/2)} \int_0^{\frac{1}{1 + R(d,p)}} \theta^{d/2 - 1}(1 - \theta)^{d/2 -1}d \theta = p
\]
if and only if
\[
\lim_{d \to \infty} \sqrt{d}\left[\frac{1}{2}- \frac{1}{1+R(d,p)}\right] = \frac{1}{\sqrt{2}}\text{erf}^{-1}(1-p),
\]
where $\text{erf}^{-1}$ is the inverse of the error function.
Straightforward manipulations show that the last relation is equivalent to 
\[
R(d,p) = 1 + \frac{2\sqrt{2}}{\sqrt{d}}\text{erf}^{-1}(1-p) + o\left(d^{-1/2}\right),
\]
or to
\begin{equation}
\label{eq:24}
\ln[R(d,p)] = \frac{2\sqrt{2}}{\sqrt{d}}\text{erf}^{-1}(1-p) + o\left(d^{-1/2}\right).
\end{equation}
We remind the reader that, in general, the notation $x_d = x + o(d^\alpha)$ means 
\[
\lim_{d \to \infty} (x_d - x) / d^\alpha = 0. 
\]
From Eqs.~(\ref{eq:23}) and (\ref{eq:24}), we learn that
\begin{equation}
\label{eq:25}
N_H(d,p) = d^{1/2} \frac{\sqrt{2}\ln(\beta_{max}/\beta_{min})}{4\text{erf}^{-1}(1-p)} + o\left(d^{1/2}\right),
\end{equation}
relation that is similar to Eq.~(9) of Ref.~\onlinecite{Fuk02}.

For condensed-phase systems, the heat capacity is usually larger than what an harmonic approximation predicts and, therefore, so is $M(\bar{\beta})$. Consequently, for such systems,
the acceptance ratio for harmonic oscillators $Ac_H(\beta, \beta')$ is usually not attained. In this conditions, all we can say is that  $N(d,p)$ must satisfy the relation
\begin{equation}
\label{eq:26}
N(d,p) \geq  d^{1/2} \frac{\sqrt{2}\ln(\beta_{max}/\beta_{min})}{4\text{erf}^{-1}(1-p)}, 
\end{equation}
inequality which demonstrates our claim that there is a curse of dimension for the parallel tempering method. In this respect, notice that running $N(d,p)$ independent replica requires $N(d,p)$ times more computational power. However, given the improvement in the quality of the sampling brought in by parallel tempering, this loss of efficiency is an acceptable price to pay.

\section{A numerical example}

In this section, we verify the validity of the empirical incomplete beta function law for the neon realization of the $\text{LJ}_{13}$ cluster. This example is representative of the type of applications one is likely to encounter in practice. We shall also illustrate the use of the incomplete beta function law for the design of optimal schedules for parallel tempering. With the help of the identity $C(\beta) = C_V(\beta) + d/2$, the empirical incomplete beta function law is expressed as a functional of the full heat capacity as 
\begin{eqnarray}
\label{2.0}
\nonumber
Ac(\beta,\beta') \approx 2 \, B\!\!\left(C(\bar{\beta})-d/2,C(\bar{\beta})-d/2\right)^{-1} \\ \times \int_0^{1/(1+R)} \theta^{C(\bar{\beta})-d/2 - 1} (1 - \theta)^{C(\bar{\beta})-d/2 -1}d \theta, 
\end{eqnarray}
where $d = 3\cdot 13 = 39$ is the dimensionality of the cluster.

The total potential energy of the $\text{Ne}_{13}$ cluster is given by
\begin{equation}
\label{2.1} V_{tot} = \sum_{i<j}^{N_p} V_{LJ}(r_{ij})+\sum_{i=1}^{N_p}
V_c(\mathbf{r_i}),
\end{equation} where $V_{LJ}(r_{ij})$ is the pair interaction of
Lennard-Jones potential
\begin{equation}
\label{2.2} V_{LJ}(r_{ij}) = 4\epsilon_{LJ}\left
        [\left( \frac{\sigma_{LJ}}{r_{ij}}\right)^{12}
       -\left( \frac{\sigma_{LJ}}{r_{ij}}\right)^{6}\right]
\end{equation} and V$_{c}(\mathbf{r_i})$ is the confining potential
\begin{equation}
\label{2.3}
V_c(\mathbf{r_i})=\epsilon_{LJ}\left(\frac{|\mathbf{r_i}-\mathbf{R_{cm}}|}{R_c}\right)^{20}.
\end{equation}
$N_p = 13$ is, of course, the number of particles in the system.  The values of the Lennard-Jones parameters
$\sigma_{LJ}$ and $\epsilon_{LJ}$ used are 2.749 {\AA} and 35.6 K,
respectively. The mass of the Ne atom was set to $m_0=20$, the rounded atomic mass of the most abundant isotope. 
$\mathbf{R_{cm}}$ is the coordinate  of the center of mass of the
cluster  and is given by
\begin{equation}
\label{2.4}
\mathbf{R_{cm}}=\frac{1}{N_p}\sum_{i=1}^{N_p} \mathbf{r_i}.
\end{equation}
 Finally, $R_c=4\sigma_{LJ}$ is the confining radius.
The role of the confining potential $V_c(\mathbf{r_i})$ is to prevent
atoms from permanently leaving the cluster since the cluster in
vacuum at any finite temperature is metastable with respect to
evaporation. 

\begin{figure}[!tbp] 
   \includegraphics[angle=0,width=7.0cm,clip=t]{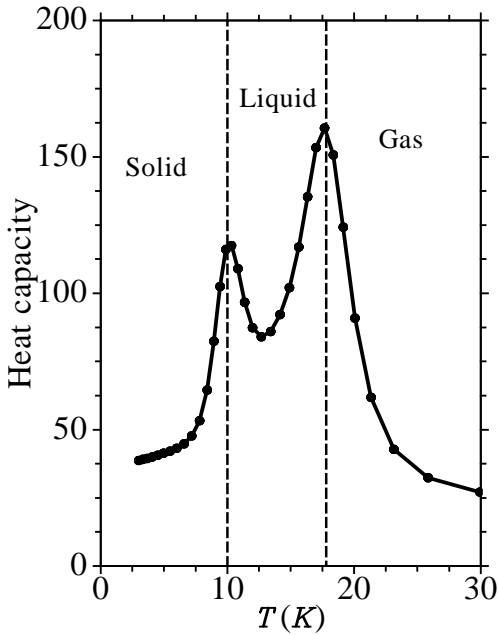} 
 \caption[sqr]
{\label{Fig:1}
Heat capacities (in units of $k_B$) as a function of the temperature $T$ (in Kelvin). The data for the plot have been computed with the help of the optimal schedule determined in Section~IV. The error bars (twice the standard deviation) are smaller than the plotting symbols. 
}
\end{figure}

The range of temperatures employed was $3~\text{K}$ to $30~\text{K}$. In this range of temperatures, the system undergoes two phase transitions: a solid-liquid transition at about $10~\text{K}$ and a liquid-gas transition at about $18~\text{K}$, respectively. Surely, the terminology we employ is more or less abuse of language, because true first-order phase transitions happen only in the limit of an infinite number of particles. However, Fig.~\ref{Fig:1} clearly shows two pronounced maxima in the heat capacity of the system, maxima that  separate the three phases. In the limit of infinite number of particles, these maxima sharpen to a delta function and their temperature value is usually lowered to the corresponding bulk values. 

The parallel tempering Monte Carlo simulations are carried out using a total of $32$ parallel streams, each running a replica of the system at a different temperature. The temperatures range from $3~\text{K}$ to $30~\text{K}$ and are distributed in geometric progression.   For each stream, the coordinate sampling is performed with the help of the Metropolis algorithm.\cite{Kal86} The basic Monte Carlo steps  consist in attempted moves of the physical coordinates  associated with a given particle. Each attempted move is then accepted or rejected according to the Metropolis logic. By attempting to move the particles one at a time, we can increase the maximum displacements and ensure better quality of sampling. The maximum displacements are adjusted in the equilibration phase of the computation, so that each of the acceptance ratios eventually lies between $40\%$ and $60\%$. The $32$ (statistically) independent streams of random numbers necessary for the simulation are obtained with the help of the Dynamic Creator package.\cite{Mat98a, Mat98b}

As a counting device, we define a \emph{pass} as the minimal set of Monte
Carlo attempts over all particles in the system. Because we update the neon atoms in successive fashion, a pass consists of $13$ basic steps. One also defines a \emph{block} as a computational unit made up of $100$ thousand passes. The size of the block is  sufficiently large  that the block averages of the various quantities computed are independent for all practical purposes. The accumulation phase of the simulation has consisted of $100$ blocks for a total of $10$ million passes per temperature. The accumulation phase has been preceded by an equilibration phase of $20$ blocks.

Parallel tempering swaps between neighboring temperatures are attempted every $100$ passes in an alternating fashion (first, with the closest lower temperature and then, with the closest higher temperature). The only exceptions are the two end temperatures, which are involved in swaps every $200$ passes, only. The acceptance probability of swaps $ac_i$ at a given temperature $\beta_i$ is computed as the fraction of accepted swaps involving that temperature. Thus, except for the end temperatures, the computed values are  equal to
\[
ac_i = \frac{1}{2}\left[Ac(\beta_{i-1}, \beta_{i}) + Ac(\beta_i, \beta_{i+1})\right].
\]

\begin{figure}[!tbp] 
   \includegraphics[angle=0,width=7.0cm,clip=t]{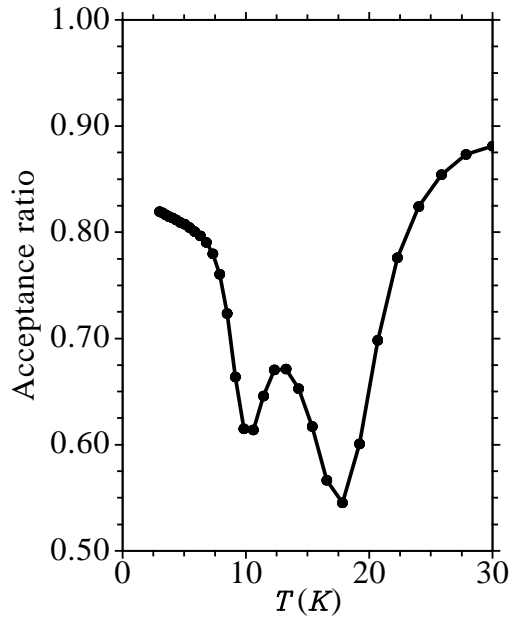} 
 \caption[sqr]
{\label{Fig:2}
Observed acceptance ratios as a function of the temperature $T$ (in Kelvin). A geometric progression schedule has been utilized. The plot is in direct relationship with the heat capacity curve (see Fig.~\ref{Fig:1}), as predicted by the incomplete beta function law. The error bars are about the size of the plotting symbols. 
}
\end{figure}

Because the intermediate temperatures are distributed in geometric progression, 
$R = \left({30}/{3}\right)^{1/31}$ is a constant. In this case, the empirical incomplete beta function law given by Eq.~(\ref{2.0}) says that the dependence of the acceptance probabilities with the temperature is a functional of the heat capacity only. This observation is well supported by the observed acceptance probabilities, which are plotted in Fig.~\ref{Fig:2}. 

\begin{figure}[!tbp] 
   \includegraphics[angle=0,width=7.0cm,clip=t]{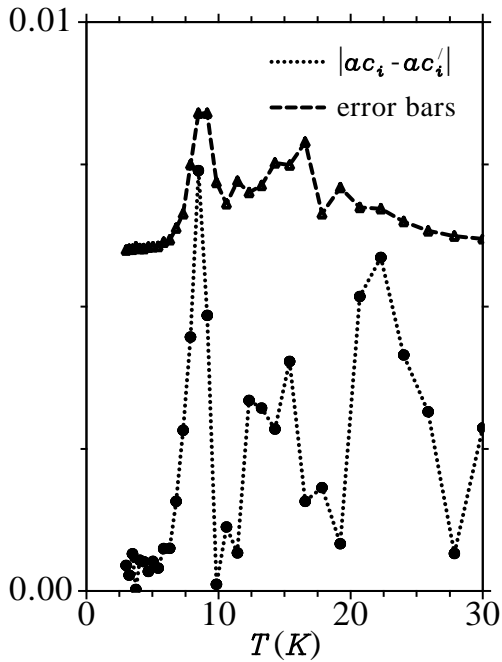} 
 \caption[sqr]
{\label{Fig:2p}
The dot line shows the absolute value of the differences between the observed values for the acceptance probability of swaps and the ones predicted by the empirical incomplete beta function law. Also plotted (dash line) are the estimated error bars for the differences $|ac_i-ac'_i|$. 
}
\end{figure}

During the Monte Carlo simulation, besides acceptance ratios,  we have also evaluated the heat capacities at different temperatures.  Then, the heat capacities have been interpolated using a cubic spline and acceptance ratios $ac'_i$ have been computed with the help of the empirical incomplete beta function law. More precisely, we have computed
\[
ac'_i = \frac{1}{2}\left[Ac'(\beta_{i-1}, \beta_{i})+Ac'(\beta_i, \beta_{i+1})\right],
\]
where $Ac'(\beta, \beta')$ is given by the right-hand side of Eq.~(\ref{2.0}). In  Fig.~\ref{Fig:2p}, we plot the absolute values of the differences between the observed and computed acceptance probabilities. These values are smaller than the estimated error bars for the same differences. In fact, the maximum difference between the two curves is less than $0.008$ and therefore, for all practical purposes, the empirical version of the incomplete beta function law is correct.

As discussed in Section~III.C, the heat capacity for solids is usually larger than what the harmonic approximation predicts, except for the low temperature limit. The liquid phase  has an even larger heat capacity, whereas the gas phase has a smaller one. In the high-temperature limit, the $C_V(\beta)$ component of the heat capacity decreases to zero. As a consequence, the acceptance probabilities increase to $1$. At the critical points, the acceptance probabilities for parallel tempering swaps decrease very much, as the heat capacities increase suddenly. This analysis  is consistent with the results shown in Fig.~\ref{Fig:2}. Therefore, the  loss of efficiency for solids and liquids is even more pronounced than the $d^{-1/2}$ decay computed for $d$-dimensional harmonic oscillators.

\begin{figure}[!tbp] 
   \includegraphics[angle=0,width=7.0cm,clip=t]{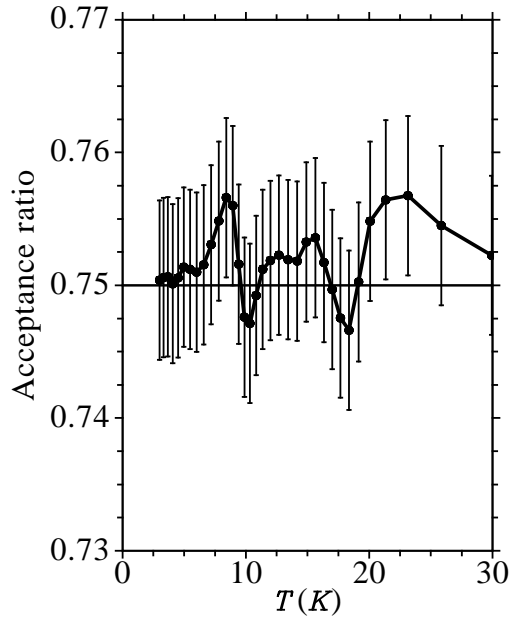} 
 \caption[sqr]
{\label{Fig:3}
Observed acceptance ratios for the optimal schedule of temperatures determined with the help of the empirical incomplete beta function law. The deviations from the ideal result of $p = 0.75$ are comparable to the statistical noise. 
}
\end{figure}

Using the strategy described in Section~III.B, we have determined the optimal schedule of temperatures necessary to achieve a constant acceptance probability of $p= 0.75$ for all swaps between neighboring temperatures. The minimal number of temperatures needed has been found to be $34$ (only the temperatures located in the interval $[3~\text{K},30~\text{K}]$ are counted). Then, we have performed a second Monte Carlo simulation to verify the validity of the schedule. The plot in Fig.~\ref{Fig:3} demonstrates that the computed schedule works very well. This explicit application illustrates the utility of the empirical incomplete beta function law for the determination of optimal temperature schedules in parallel tempering simulations. 

\section{Summary and conclusions}

We have successfully and rigorously related the acceptance probabilities for parallel tempering swaps to the ratio between the temperatures involved in the swap and the average statistical fluctuations of the potential at some intermediate temperature. The respective law, called the incomplete beta function law, is exact for harmonic oscillators and of order $O(|\beta'-\beta|^3)$ for arbitrary systems. We have also demonstrated that there is a loss of efficiency in parallel tempering simulations of condensed-phase systems with the increase of dimensionality. The loss of efficiency is at least $d^{-1/2}$, the value computed for harmonic oscillators. 

Motivated by the fact that the existent Monte Carlo codes do not allow for the computation of the average potential fluctuations without extensive reprogramming, we have developed and tested the empirical incomplete beta function law. This empirical law connects the acceptance probabilities of the parallel tempering swaps with the heat capacity of the system. The law has been extensively verified for the $\text{LJ}_{13}$ cluster, on a range of temperatures that spanned three thermodynamic phases. The empirical incomplete beta function law provides a direct justification of the observation that a geometric progression is the optimal schedule for systems and regions of temperatures where the heat capacity is almost constant. Finally, the use of the empirical incomplete beta function law for the construction of optimal temperature schedules has been demonstrated. 

As opposed to its empirical version, the incomplete beta function law given by Theorem~1 is an exact mathematical statement,  valid for all systems asymptotically, for close enough temperatures. For strongly anharmonic systems (for instance, systems for which the sampling is performed on a lattice, such as spin glasses and self-avoiding random walks), the empirical version of the incomplete beta function law may fail. In such cases, the rigorous incomplete beta function law should be used for the development of optimal temperature schedules. As discussed in the text, the evaluation of the average
\begin{eqnarray}
\label{eq:3a}
M(\beta)=  \frac{1}{Q(\beta)^2}\int_{\mathbb{R}^d}d \mathbf{x} \int_{\mathbb{R}^d}d \mathbf{x}' e^{-{\beta} V(\mathbf{x})}e^{-{\beta} V(\mathbf{x}')}\nonumber \\ \times \beta \left|V(\mathbf{x}')-V(\mathbf{x})\right| \quad
\end{eqnarray}
requires however a double integral over the configuration space. This integral can be computed by doubling the number of temperatures in the parallel tempering schedule, as follows
\[
\beta_0 = \beta_1 < \beta_2 = \beta_3 < \beta_4 = \beta_5 < \cdots < \beta_{2N-2} = \beta_{2N-1}.
\]
Then, to compute $M(\beta_i)$, one collects the values of the differences $\beta_i\left|V(\mathbf{x})-V(\mathbf{x}')\right|$ any time a swap between equal neighboring temperatures $\beta_i$ and $\beta_{i\pm 1}$ is attempted. Of course, swaps between equal temperatures are always accepted. The values $M(\beta_i)$ are interpolated by a cubic spline. The determination of the optimal temperature schedule then proceeds in a way similar to the approach utilized in Section~IV.

While the reader may object that the introduction of an intercalated set of identical temperatures is an additional computational burden, many times, there are certain advantages in doing so. For large dimensional systems, coupled independent replica running at identical temperatures constitute an elegant device for parallelizing the Monte Carlo code, whenever the number of available compute nodes is at least twice the number of temperatures in the parallel tempering schedule. Nowadays, with the advent of inexpensive cluster computing, this is  the case with many research groups. Furthermore,  in this setting, the computation of the heat capacity and other properties of the system can be done by means of unbiased estimators, as shown by Eq.~(\ref{eq:18a}).

On a more general level, we hope that a better understanding of the laws governing the acceptance probabilities for swaps in parallel tempering methods  may lead to useful research in improving the efficiency of the methods. In the meantime, we recommend that the dimensionality of the systems be maintained as low as possible, for instance, by adiabatically reducing those degrees of freedom that do not lead to significant degradation in the quality of the final results.

\begin{acknowledgments} The first author acknowledges support from the National
Science Foundation through awards No. CHE-0095053 and CHE-0131114. CVC is supported by the MRSEC program at Brown University. The authors
also wish to thank Professors J. D. Doll and D. L. Freeman for useful suggestions and interesting discussions on the subject. 
\end{acknowledgments}

\appendix
\section{Evaluation of integrals}

For an arbitrary $d$-dimensional harmonic oscillator whose global minimum is zero, the density of states is given by the formula $\Omega(U) = (2\pi)^{d/2}U^{d/2 - 1}/\Gamma(d/2)$. In these conditions, it is but a simple exercise to show that the acceptance probability for the parallel tempering swaps is given by the equation
\begin{eqnarray}
\label{eq:A1}  \nonumber
Ac_H(\beta,\beta') =\frac{(\beta\beta')^{d/2}}{\Gamma(d/2)^2} \int_0^\infty \int_0^\infty e^{-\beta U}e^{-\beta' U'}  U^{d/2-1}\\ \times  U'^{d/2-1} \min\left\{1, e^{(\beta'-\beta)(U'-U)}\right\}dU dU'. \quad
\end{eqnarray} 
We also want to evaluate the quantity $M_H(\beta)$ [see Eq.~(\ref{eq:15})], which is given by the formula
\begin{eqnarray}
\label{eq:A2}  \nonumber
M_H(\beta) =\frac{\beta^{d+1}}{\Gamma(d/2)^2} \int_0^\infty \int_0^\infty e^{-\beta U}e^{-\beta U'}  U^{d/2-1}\\ \times  U'^{d/2-1} |U'-U| dU dU'. \quad
\end{eqnarray} 
While for an harmonic oscillator the parameter $d$ is an integer, we compute the two integrals above under the assumption that $d$ is a strictly positive real number. 
We prove that the value of $Ac_H(\beta, \beta')$ is given by the incomplete beta function
\begin{equation}
\label{eq:A1p}
Ac_H(\beta, \beta') = \frac{2}{B(d/2,d/2)} \int_0^{1/(1 + R)} \theta^{d/2 - 1}(1 - \theta)^{d/2 -1}d \theta,
\end{equation}
where $B(a, b)$ is Euler's beta function
\[
B(a,b)= \int_0^{1} \theta^{a - 1}(1 - \theta)^{b -1}d \theta
\]
 and
\[
R = \max\left\{\frac{\beta'}{\beta}, \frac{\beta}{\beta'}\right\}.
\]
In addition, we prove that 
\begin{equation}
\label{eq:A2p}
M_H(\beta) = 2^{2-d}B(d/2,d/2)^{-1}.
\end{equation}

Because the function $Ac_H(\beta, \beta')$ is symmetrical in its arguments, we may assume without loss of generality that $\beta' \geq \beta$, so that $R = \beta'/\beta \geq 1$. 
By decomposing the domain of the integral against $U'$ in Eq.~(\ref{eq:A1})  in two regions with $U' < U$ and $U' \geq U$ respectively, it is straightforward to see that
\begin{equation}
\label{eq:A3}
Ac_H(\beta, \beta') = 1 - I(\beta, \beta') + I(\beta', \beta),
\end{equation}
where
\begin{eqnarray}
\label{eq:A4}  \nonumber
I(\beta,\beta') =\frac{(\beta\beta')^{d/2}}{\Gamma(d/2)^2} \int_0^\infty dU \int_0^U dU' e^{-\beta U}\\ \times e^{-\beta' U'} U^{d/2-1}U'^{d/2-1}.
\end{eqnarray} 
Performing the substitution $U' = U \theta$, we obtain
\begin{eqnarray}
\label{eq:A5}  \nonumber
I(\beta,\beta') =\frac{\beta^d R^{d/2}}{\Gamma(d/2)^2} \int_0^\infty dU \int_0^1 d
\theta e^{-\beta(1+R\theta) U}\\ \times U^{d-1} \theta^{d/2-1}  = \frac{\Gamma(d) R^{d/2}}{\Gamma(d/2)^2}\int_0^1 d
\theta \frac{\theta^{d/2-1} }{(1+R\theta)^d}. 
\end{eqnarray} 
Performing the change of variables $\theta = R^{-1}(t^{-1}-1)$, we conclude
\begin{eqnarray}
\label{eq:A6}\nonumber
I(\beta, \beta') = 
\frac{1}{B(d/2,d/2)} \int_{1/(R+1)}^{1} t^{d/2-1}(1-t)^{d/2-1}dt \\ = 1 -
\frac{1}{B(d/2,d/2)} \int_{0}^{1/(R+1)} t^{d/2-1}(1-t)^{d/2-1}dt, \quad
\end{eqnarray}
where 
\[
B(d/2,d/2) = \frac{\Gamma(d/2)^2}{\Gamma(d)} = \int_{0}^1 t^{d/2-1}(1-t)^{d/2-1}dt
\]
is Euler's beta function. The value of $I(\beta', \beta)$ is obtained by replacing $R$ with $1/R$ in the first expression of Eq.~(\ref{eq:A6}). We compute 
\begin{equation}
\label{eq:A7}\nonumber
I(\beta', \beta) = \frac{1}{B(d/2,d/2)} \int_{0}^{1/(1+R)} t^{d/2-1}(1-t)^{d/2-1}dt.
\end{equation}
Eq.~(\ref{eq:A1p}) follows easily from Eqs.~(\ref{eq:A3}), (\ref{eq:A6}), and (\ref{eq:A7}), after easy simplifications. 

If applied to  Eq.~(\ref{eq:A2}), the same decomposition and coordinate transformations used in the proof of Eq.~(\ref{eq:A1p}) lead to the equation
\begin{eqnarray}
\label{eq:A8}\nonumber
M_H(\beta) = 2\frac{\Gamma(d+1)}{\Gamma(d/2)^2}\int_{1/2}^1(2\theta -1)\theta^{d/2-1}(1-\theta)^{d/2-1} d\theta\\ = 2dB(d/2,d/2)^{-1}\left[2I_1 - \frac{1}{2}B(d/2,d/2)\right], \qquad
\end{eqnarray}
where
\[
I_1 = \int_{1/2}^1 \theta^{d/2}(1-\theta)^{d/2-1}d\theta.
\]
Integrating by parts the last integral, we obtain
\begin{equation}
\label{eq:A9}
I_1 =\frac{2}{d2^d} + \int_{1/2}^1 \theta^{d/2-1}(1-\theta)^{d/2}d\theta  = 
\frac{2}{d2^d} + I_2, 
\end{equation}
where
\[
I_2 = \int_{0}^{1/2}(1-\theta)^{d/2-1} \theta^{d/2}d\theta.
\]
Combining Eq.~(\ref{eq:A9}) with the identity
\[I_1 + I_2 = B(d/2,d/2+1) = \frac{1}{2}B(d/2,d/2),\]
we obtain
\[
2I_1 = \frac{2}{d2^d}  + \frac{1}{2}B(d/2,d/2),
\]
which, after replacement in Eq.~(\ref{eq:A8}), produces Eq.~(\ref{eq:A2p}).

\section{A limiting theorem}
\begin{A2}
Let $\{\alpha_d\}_{d\geq 1}$ be a sequence of positive numbers convergent to $\alpha > 0$. Then,
\begin{eqnarray}
\label{eq:B1} \nonumber
\lim_{d \to \infty} \frac{2}{B(d/2,d/2)} \int_0^{\frac{1}{2}-\frac{\alpha_d}{d^{1/2}}} \theta^{d/2 - 1}(1-\theta)^{d/2 - 1}d\theta \\ = 
1-\textrm{erf}\,(2^{1/2}\alpha), \quad 
\end{eqnarray}
where 
\[
\textrm{erf}\,(x) = \frac{2}{\sqrt{\pi}}\int_0^x e^{-t^2}dt
\]
is the error function. 
\end{A2}

\emph{Proof.} We compute
\begin{eqnarray*}&&
I_d = \frac{2}{B(d/2,d/2)} \int_0^{\frac{1}{2}-\frac{\alpha_d}{d^{1/2}}} \theta^{d/2 - 1}(1-\theta)^{d/2 - 1}d\theta \\&& = 1 - \frac{2}{B(d/2,d/2)} \int_{\frac{1}{2}-\frac{\alpha_d}{d^{1/2}}}^{1/2} \theta^{d/2 - 1}(1-\theta)^{d/2 - 1}d\theta \\
&& = 1- \frac{2}{2^{d-2}B(d/2,d/2)} \int_0^{\frac{\alpha_d}{d^{1/2}}} (1-4t^2)^{d/2 - 1}dt,
\end{eqnarray*}
where we have used the transformation of coordinates $\theta =  1/2 - t$ for the last expression. Next, we perform the substitution $t = \theta \alpha_d / d^{1/2}$ and obtain 
\begin{eqnarray*}
I_d = 1 - \frac{4\alpha_d}{d^{1/2}2^{d-1}B(d/2,d/2)} \int_0^{1} \left(1-\frac{4\theta^2\alpha_d^2}{d}\right)^{d/2 - 1}d\theta
\end{eqnarray*}

Sterling's formula implies 
\begin{eqnarray*}&&
\lim_{d \to \infty} \frac{1}{B(d/2,d/2)2^{d-1}d^{1/2}} \\ &&= \lim_{d \to \infty} \frac{\Gamma(d)}{\Gamma(d/2)^2} \frac{1}{2^{d-1}d^{1/2}} = \frac{1}{\sqrt{2\pi}}.
\end{eqnarray*}
Therefore, remembering that $\alpha_n \to \alpha$, we obtain
\begin{eqnarray*}
\lim_{d \to \infty} I_d = 1- \frac{4\alpha}{\sqrt{2\pi}} \lim_{d \to \infty} \int_0^{1} \left(1-\frac{2\theta^2\alpha_d^2}{d/2}\right)^{d/2 - 1}d\theta
\end{eqnarray*}
The equality 
\[
\lim_{d \to \infty} \left(1-\frac{2\theta^2\alpha_d^2}{d/2}\right)^{d/2 - 1} = e^{-2 \theta^2 \alpha^2},
\]
the fact that the above sequences are bounded by $1$ for all $\theta \in [0,1]$, and the dominated convergence theorem imply
\[
\lim_{d \to \infty}\int_0^1 \left(1-\frac{2\theta^2\alpha_d^2}{d/2}\right)^{d/2 - 1} d\theta= \int_0^1 e^{-2 \theta^2 \alpha^2}d\theta.
\]
Thus, 
\begin{eqnarray*}&&
\lim_{d \to \infty} I_d = 1- \frac{4\alpha}{\sqrt{2\pi}} \int_0^1 e^{-2 \alpha^2 \theta^2} d \theta \\ && = 1 - \frac{2}{\sqrt{\pi}} \int_0^{2^{1/2}\alpha} e^{-t^2} d t = 1 - \textrm{erf}\,(2^{1/2}\alpha)
\end{eqnarray*}
and the theorem is proven. \hspace{\stretch{1}} $\Box$



\end{document}